\documentclass{jocg}
\usepackage{amsfonts}
\usepackage{amsthm}
\usepackage{graphicx}
\usepackage{cite}

\title{%
\MakeUppercase{Happy Endings for Flip Graphs}
\thanks{This work was supported in part by NSF grant
0830403 and by the Office of Naval Research under grant
N00014-08-1-1015.}
}

\author{David Eppstein
\thanks{\affil{Computer Science Department, Donald Bren School of Information and Computer Sciences, University of California, Irvine}, \email{eppstein@uci.edu}}}

\DeclareSymbolFont{AMSb}{U}{msb}{m}{n}
\DeclareSymbolFontAlphabet{\Bbb}{AMSb}

\def\hull{\mathop{\mbox{\small\rm CH}}}
\def\delaunay{\mathop{\mbox{\small\rm DT}}}
\def\quadgraph{\mathop{\mbox{\small\rm QG}}}
\def\flipgraph{\mathop{\mbox{\small\rm FG}}}

\newtheorem{theorem}{Theorem}
\newtheorem{lemma}{Lemma}

\begin{document}
\maketitle   

\begin{abstract}
We show that the triangulations of a finite point set form a flip graph that can be embedded isometrically into a hypercube, if and only if the point set has no empty convex pentagon. Point sets with no empty pentagon include intersections of lattices with convex sets, points on two lines, and several other infinite families. As a consequence, flip distance in such point sets can be computed efficiently.
\end{abstract}

\section{Introduction}

\begin{figure}[t]
\centering\includegraphics[width=5in]{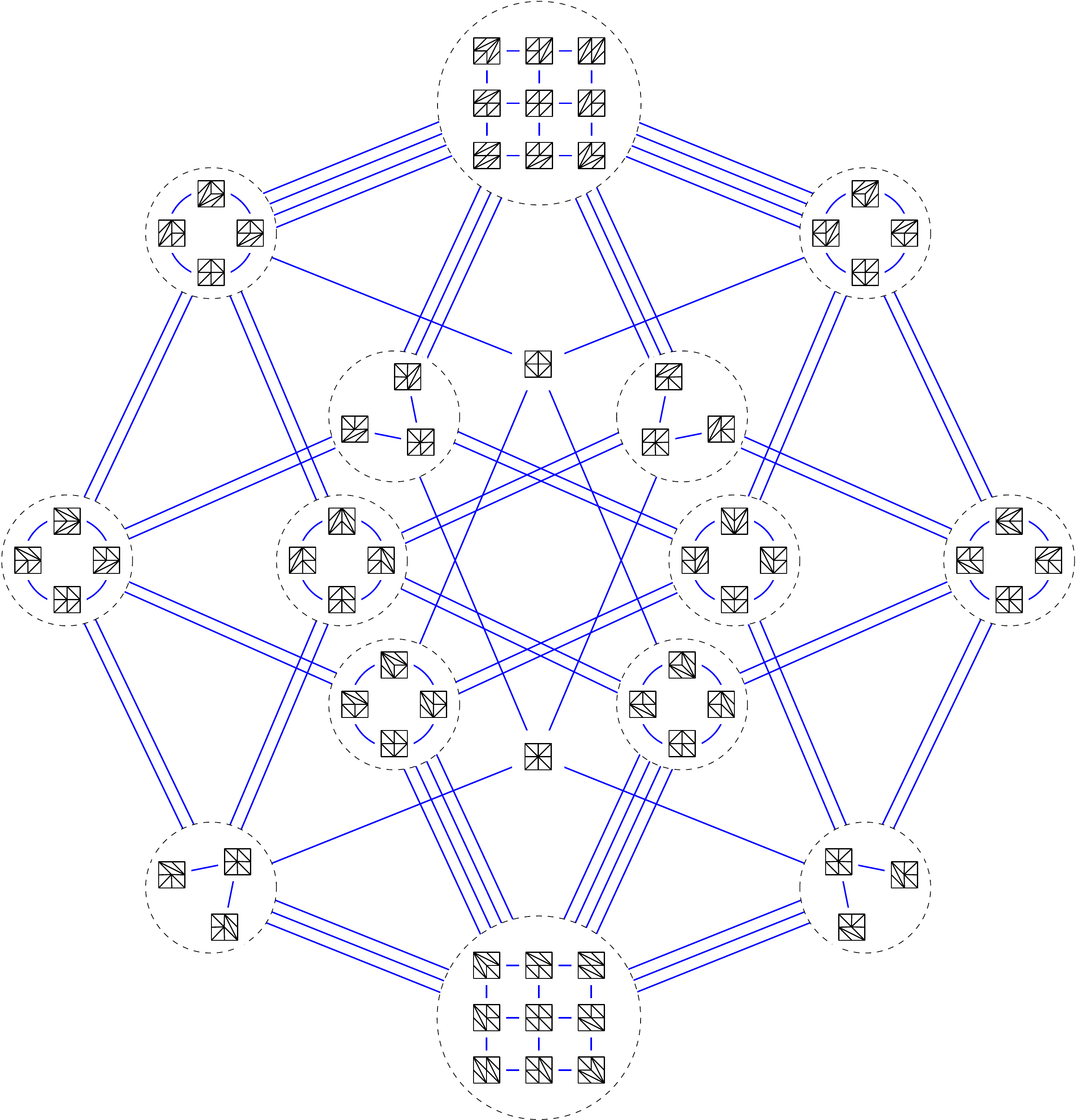}
\caption{A high-level view of the flip graph of a $3\times 3$ grid of points. The 64 triangulations of this point set are grouped so that each flip within a group adds or removes an edge between adjacent grid points and the flips between two groups involve longer edges. Within each group, the flips are shown directly as edges connecting pairs of triangulations, while the number of lines between each pair of groups indicates the number of flip graph edges connecting pairs of triangulations in those two groups.}
\label{fig:3x3flipgroups}
\end{figure}

\emph{Triangulations}, partitions of the convex hull of a point set into triangles having the given points as vertices and meeting edge-to-edge, form a fundamental object of study in computational geometry, of importance due to applications including computer graphics and finite element mesh generation.
A point set may have many different triangulations, and there are many known algorithms for finding high-quality triangulations using various optimization criteria~\cite{BerEpp-CEG-92}.
Flips, flip distance, and the flip graph (all defined below) form important tools in organizing the set of possible triangulations, and in performing local optimization heuristics for finding good triangulations.

A \emph{flip} in a triangulation is the replacement of two triangles that share an edge by the other two triangles that cover the same quadrilateral. Flipping has long been important in the study of triangulations, both topologically since the discovery that any two triangulations of the same topological surface can be connected to each other by flips~\cite{Wag-JDMV-36} and computationally from the fact that repeated flips can be used to compute Delaunay triangulations~\cite{Law-72,Sib-CJ-73}. However, despite this long study, much about flipping remains mysterious. In particular, it is not known how to compute efficiently the \emph{flip distance} between triangulations, the minimum number of flips needed to transform one triangulation into another. Even for convex point sets, it remains unknown whether flip distances may be computed in polynomial time, although in this case tight linear bounds are known on how large the flip distance can be as a function of the number of points~\cite{HurNoy-CGTA-99,SleTarThu-JAMS-88}. For nonconvex points, flip distances may grow superlinearly compared to the number of points~\cite{BosHur-CGTA-09,HanOttSch-JUCS-96,HurNoyUrr-DCG-99} and it is unknown how to approximate this number efficiently and accurately. The higher dimensional analogue of flip distance is not, in general, well-defined: there exist pairs of triangulations that cannot be converted into each other by flips~\cite{San-JAMS-00}.

The set of triangulations of a planar point set, and the flip distances between any two triangulations, form a \emph{metric space}: flip distances are symmetric, nonnegative, and satisfy the triangle inequality.
The metric space of triangulations and flip distances can be represented combinatorially as an undirected graph, the \emph{flip graph}, that has a vertex per triangulation and an edge per flip between two triangulations (for examples, see Figures~\ref{fig:3x3flipgroups} and~\ref{fig:fg6}). The flip distance between any two triangulations is equal to the length of the shortest path between the corresponding vertices in this graph. However, except for very small numbers of points it is not feasible to compute flip distances by constructing the flip graph explicitly and applying a graph shortest path algorithm: the number of triangulations may be exponential in the number of points of the input point set (in fact, for points in general position, the number of triangles is always exponential, due to the existence of linearly many independently flippable edges~\cite{HurNoyUrr-DCG-99}) so the flip graph is generally too large to construct.

\begin{figure}
\centering\includegraphics[width=3.25in]{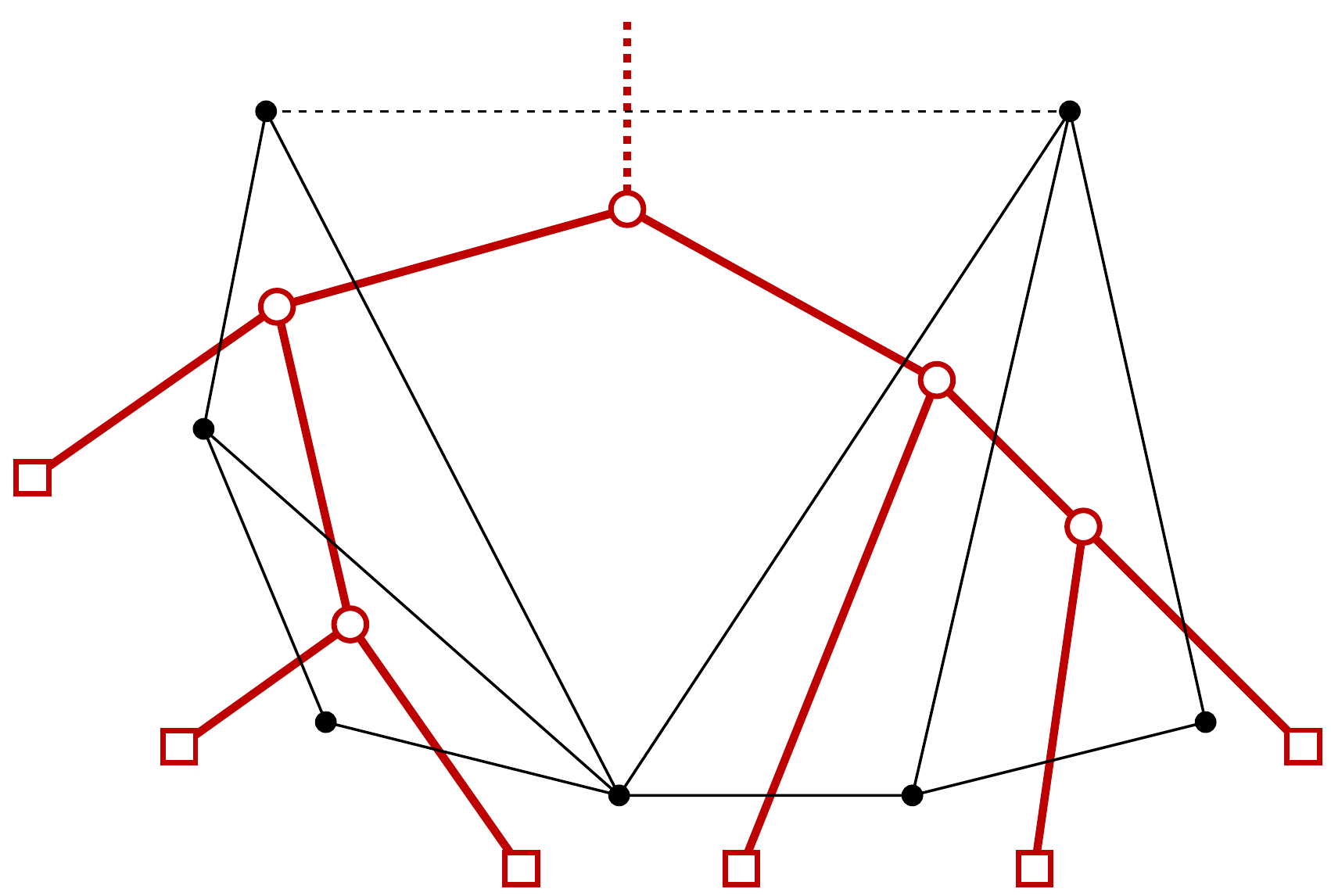}
\caption{A triangulation of a point set in convex position (with a designated root edge of the convex hull shown dashed) and its dual binary tree (with a dashed vertical line segment connecting to the root vertex).}
\label{fig:binary-tree-dual}
\end{figure}

There are also strong connections between flip graphs, rotation in binary trees, and polyhedral combinatorics.
If a set of $n$ points is in \emph{convex position} (its points form the vertices of a convex polygon, with one polygon edge chosen arbitrarily as root), then its triangulations are in one-to-one correspondence, by a form of planar duality, to the binary trees with $n-1$ leaves~\cite{SleTarThu-JAMS-88}; see Figure~\ref{fig:binary-tree-dual} for an example.
According to this correspondence, a flip of a triangulation corresponds to a binary tree rotation, a standard operation in the theory of data structures that interchanges the heights of two adjacent tree nodes while preserving the left-to-right traversal ordering of the tree nodes~\cite{SleTarThu-JAMS-88}.
From this point of view, the flip graphs of convex point sets have been shown to be skeletons of polytopes, called \emph{associahedra} or \emph{Stasheff polytopes} (Figure~\ref{fig:fg6})~\cite{Sta-TAMS-63,Tam-NAW-62}. Associahedra are examples of \emph{secondary polytopes}~\cite{BilFilStu-AM-90,GelKapZel-SMD-89}, which can be defined for arbitrary point sets without restricting the points to be in convex position. However the vertices of the secondary polytope describe only the \emph{regular triangulations}, triangulations that can be formed by orthogonal projection of a higher dimensional convex polyhedron. For some sets of points that are not in convex position, there exist non-regular triangulations. Therefore, in such cases, the vertices and edges of the secondary polytope do not represent the entire flip graph.

In this paper we study flip distances and flip graphs of highly nonconvex point sets, those in which no five points of the set form the vertices of an empty pentagon. Rabinowitz~\cite{Rab-Geo-05} has called this property of having no empty pentagon the ``pentagon property''. It is known (a relative of the famous Happy Ending problem of Erd\H{o}s and Szekeres~\cite{ErdSze-CM-35}) that sets of ten or more points in general position must always have an empty pentagon~\cite{Har-EM-78}, but our results apply to point sets not necessarily in general position; for instance, the intersection of the integer lattice with any bounded convex subset of the plane forms a set with no empty pentagon.  We show that the flip graph of any point set with no empty pentagon is a \emph{partial cube}, a graph that can be embedded isometrically into a hypercube~\cite{Djo-JCTB-73}. Conversely any point set with an empty pentagon cannot have a flip graph that is a partial cube. Via this characterization, flip distances for triangulations of point sets without empty pentagons may be calculated easily and efficiently.

Emo Welzl, in an invited talk at Graph Drawing 2006~\cite{Wel-GD-06},\footnote{See also Kaibel and Ziegler~\cite{KaiZie-BCS-03}.} claimed that, for any $n\times n$ grid of points in the plane, the flip graph of the grid can be represented (not necessarily isometrically) as an induced subgraph of a hypercube. This is a special case of our result, and was the original inspiration for the investigations reported herein.

Subsequent to the preliminary conference publication of this paper~\cite{Epp-SCG-07}, Abel \emph{et al.} have also investigated point sets with the empty pentagon property: as they show, sufficiently large point sets with the empty pentagon property have arbitrarily large subsets of collinear points~\cite{AbeBalBos-09}. The analogous result for points with no empty quadrilateral is an immediate consequence of our Theorem~\ref{thm:quad-equivalence}.

\section{New results}
\label{sec:new-results}

We prove the following results:
\begin{itemize}
\item A finite point set has exactly one triangulation if and only if it has no empty quadrilateral. Moreover, the finite point sets having this property are exactly the vertex sets of planar graphs with dilation one; in previous work we completely classified these sets.

\item The flip graph of a finite point set forms a partial cube if and only if the point set has no empty pentagon. Moreover, the point sets having this property are exactly those for which another associated geometric graph, the \emph{quadrilateral graph}, is a forest.

\item If a set of $n$ points has no empty pentagon, it has $O(n^2)$ empty quadrilaterals. There exist point sets for which this bound is tight.

\item We can find an empty pentagon in a set of $n$ points, if one exists, or if there is no empty pentagon construct the quadrilateral graph, in time $O(n^2)$.

\item If a set of $n$ points has no empty pentagon, then we may compute the flip distance between any two triangulations of the point set in time $O(n^2)$.

\item We provide a polynomial-time-computable lower bound on the flip distance between triangulations in any finite point set. When the point set has no empty pentagon, or consists solely of the vertices of a pentagon, this bound is exactly equal to the flip distance. However this bound is inaccurate for any point set containing an empty hexagon. It would be of interest to characterize more precisely the point sets on which this bound is tight.
\end{itemize}

In addition we provide several new examples of finite point sets that do not have an empty pentagon. They include: any intersection of a lattice with a bounded convex set, any finite set of points drawn from the union of two lines, any finite set of points drawn from the union of three rays with coincident apexes and no concave angle, several other infinite families of examples, and several sporadic examples.

\begin{figure}[t]
\centering\includegraphics[width=5.75in]{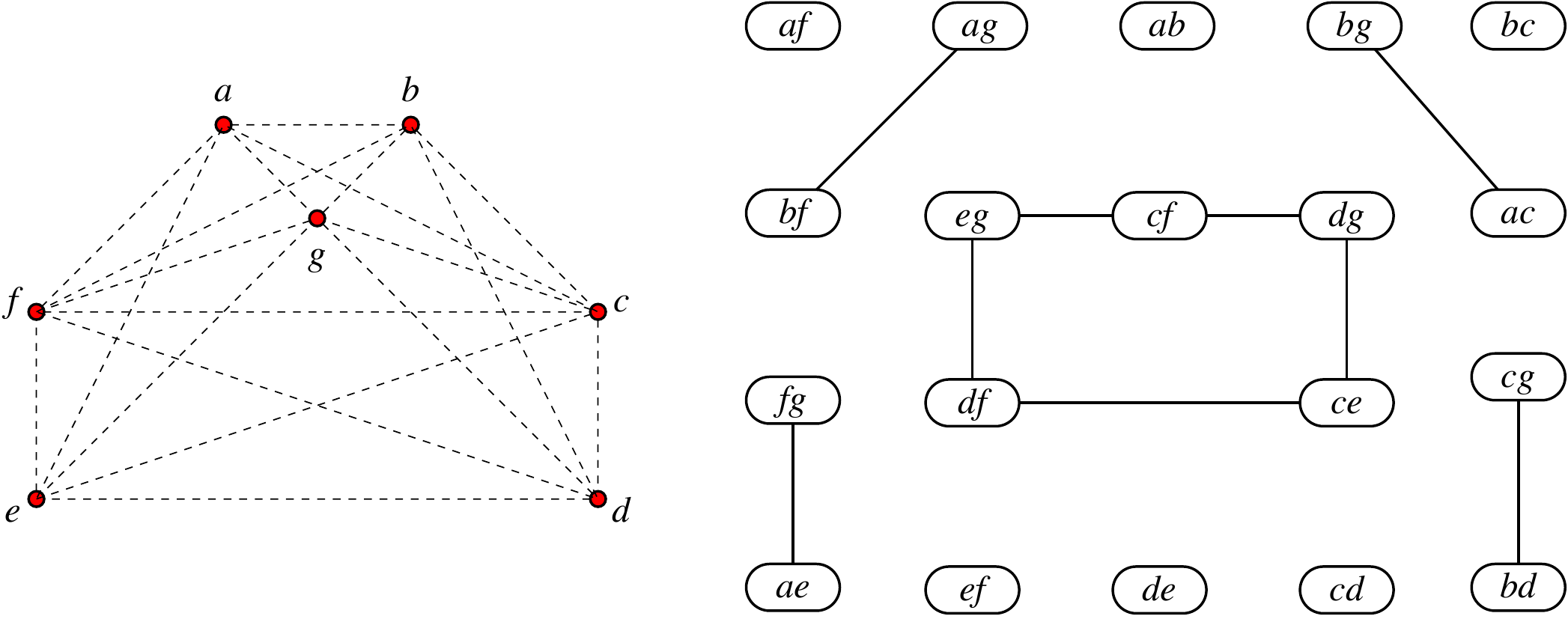}
\caption{A point set and its quadrilateral graph.}
\label{fig:quadgraph}
\end{figure}

\section{Geometric definitions}

For any point set $P$, let $\hull P$ denote the (closed) convex hull of $P$. If $\hull (P\setminus\{a\})\ne\hull P$ we say that $a$ is a \emph{vertex} of the hull.
Let $|P|$ denote the cardinality of $P$. We say that $P$ contains an \emph{empty $k$-gon} if there exists a subset $S$ of $P$ with $|S|=k$ such that $P\cap\hull S=S$ and such that each point of $S$ is a vertex of $\hull S$. That is, $S$ forms the vertices of a convex polygon which does not include any other points of $P$. In particular we say that $P$ contains an \emph{empty quadrilateral} when there exists a set $S$ with these properties and with $k=|S|=4$, and we say that $P$ contains an \emph{empty pentagon} when  there exists a set $S$ with these properties and with $k=|S|=5$. Note that three collinear points may never belong to an empty $k$-gon, because the middle point of the three belongs to the convex hull but is not a vertex.

We define a \emph{diagonal} of $P$ to be a line segment $ab$ such that $P\cap\hull\{a,b\}=\{a,b\}$;
that is, in the notation above, it is an empty 2-gon. Equivalently, a diagonal is an edge in the \emph{visibility graph} of~$P$. A \emph{triangulation} of $P$ is any graph having as its vertices all the points of $P$ and as its edges a maximal set of diagonals that do not cross each other.  Removing the edges and vertices of the triangulation from the plane partitions the remaining points of the plane into connected components that we call \emph{faces}. For a maximal set of diagonals, there is one unbounded face (the set-theoretic complement of the convex hull of $P$) and all remaining faces must be triangles; see, e.g., \cite[Lemma~1]{BerEpp-CEG-92}.

In addition, we define the following geometric graphs for a point set $P$:
\begin{itemize}
\item Let $\delaunay P$ denote the Delaunay triangulation of $P$, in which an edge is present if and only if there exists a closed circle containing no other points of $P$. We will only consider Delaunay triangulations of point sets in which no four points are cocircular, so the Delaunay triangulation is uniquely defined.
\item We define the \emph{quadrilateral graph} $\quadgraph P$ to be a graph having a vertex for each diagonal of $P$, and an edge connecting diagonals $ab$ and $cd$ whenever line segments $ab$ and $cd$ cross and $\{a,b,c,d\}$ form the vertices of an empty quadrilateral (Figure~\ref{fig:quadgraph}).
\item Let $\flipgraph P$ denote the \emph{flip graph} of $P$, a graph with a vertex for each triangulation of $P$ and an edge connecting two triangulations that differ by a single \emph{flip}: the removal of one diagonal and its replacement by another diagonal. Observe that the two diagonals involved in a flip must necessarily be connected by an edge in the quadrilateral graph.
\end{itemize}

The \emph{flip distance} between any two triangulations is defined to be the minimum number of flips needed to transform one triangulation into the other, or equivalently the unweighted distance between the corresponding vertices in the flip graph.

A \emph{Delaunay flip} of a diagonal $ac$ is a diagonal $bd$ such that $bd$ crosses $ac$ and such that $bd$ belongs to $\delaunay \{a,b,c,d\}$. Any triangulation that is not Delaunay can be transformed into the Delaunay triangulation by replacing diagonals according to a sequence of Delaunay flips~\cite{Law-72,Sib-CJ-73}. 
In any point set, each diagonal $ac$ that is not itself part of the Delaunay triangulation has at least one Delaunay flip; for instance, if $b$ and $d$ are chosen to be the points on either side of line $ac$ such that angles $abc$ and $adc$ are as large as possible, then replacing $ac$ by $bd$ is a Delaunay flip.

\section{Partial cubes}

An \emph{isometric embedding} of a graph $G$ into another graph $H$ is a function~$f$ from the vertices of $G$ to the vertices of $H$ that preserves distances: if $u$ and $v$ are any two vertices in $G$, then $d_G(u,v)=d_H(f(u),f(v))$. If $G$ is isometrically embedded into $H$, it must form an induced subgraph of $H$, but not every induced subgraph is isometrically embedded.
A \emph{partial cube} is a graph that can be embedded isometrically into a hypercube. Another way of stating this is that, if $G$ is a partial cube, we can label each vertex of $G$ with a bitvector, in such a way that unweighted distance in $G$ equals Hamming distance of labels. Thus, a partial cube structure on a graph enables us to look up distances between vertices easily; see e.g. Chepoi \emph{et al.}~\cite{CheDraVax-SODA-02} for data structures using partial cube embeddings to compute distances in certain graph families. There are many applications and examples of partial cubes~\cite{EppFalOvc-07}, and a mathematical characterization of these graphs due to Djokovi{\'c}~\cite{Djo-JCTB-73} allows them to be recognized efficiently~\cite{Epp-SODA-08}.

Although much is known about partial cubes, for the purposes of this paper we need only some simple observations about them:

\begin{itemize}
\item Any partial cube is bipartite. This follows from the facts that a partial cube is isomorphic to an induced subgraph of a hypercube and that every hypercube is bipartite. Alternatively, one can derive a bipartition from the parity of the number of nonzero bits in the bitvectors labeling the vertices of a partial cube.
\item Any tree is a partial cube.  To find a partial cube labelling of a tree, choose a root, label each vertex $v$ by a vector of bits, one bit per edge, and set the bit value for edge $e$ in $v$'s label to 1 if $e$ belongs to the path from $v$ to the root and 0 otherwise.
\item The Cartesian product of partial cubes is another partial cube. The labels for vertices in the product can be formed by concatenating the labels from the factors of the product.
\item If $G$ is a partial cube, and $G'$ is embedded isometrically into $G$, then $G'$ is also a partial cube. The labels for the vertices of $G'$ can be copied from the labels of their images in $G$.
\end{itemize}

\section{Quadrilaterals, dilation, and unique triangulation}
\label{sec:no-empty-quad}

As a warmup to our main result (Theorem~\ref{thm:pent-equivalence}) describing several equivalent characterizations for point sets with no empty pentagon, we prove a similar equivalence of characterizations for point sets with no empty quadrilateral.

\begin{figure}[t]
\centering\includegraphics[width=1.75in]{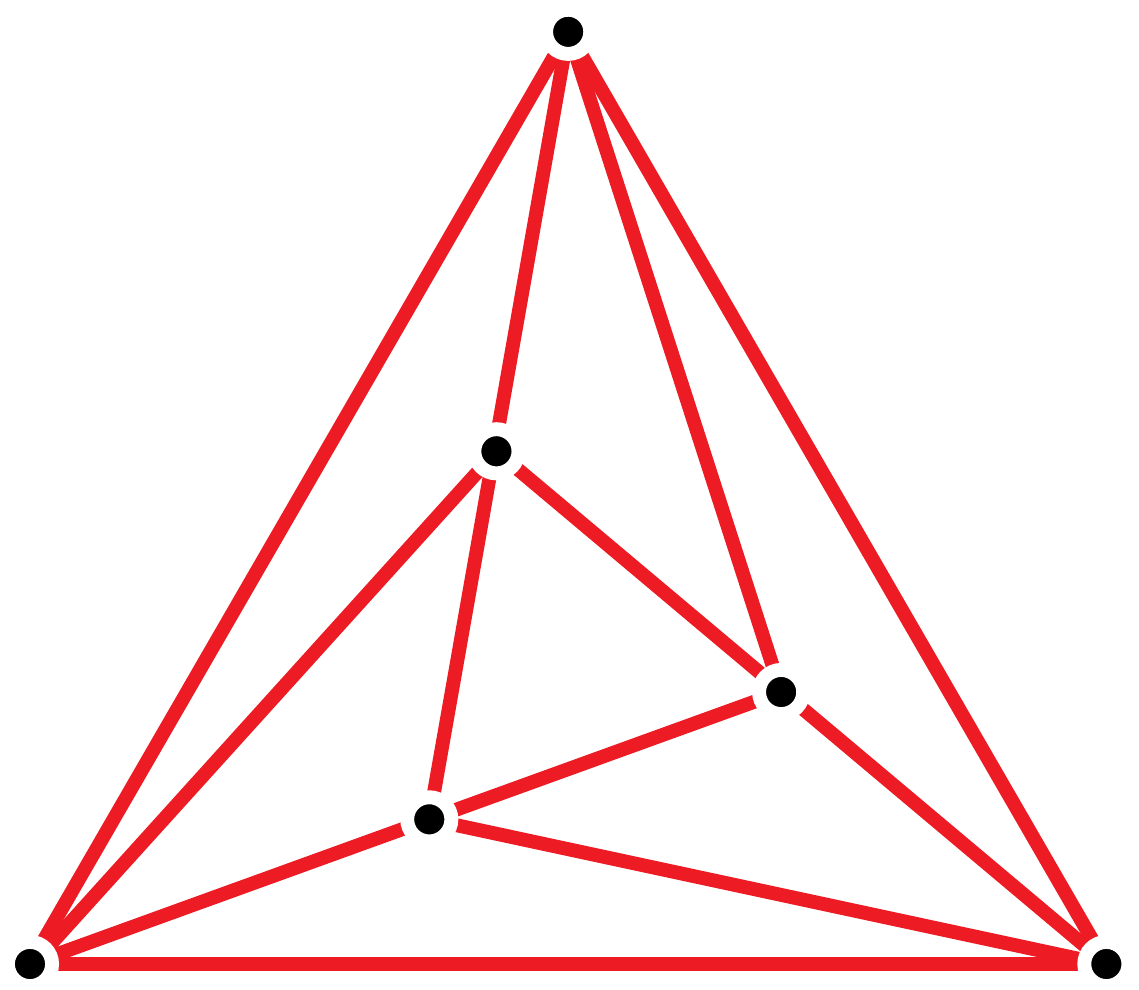}
\caption{A planar graph on six vertices with dilation one.}
\label{fig:dilation-free}
\end{figure}

If $G$ is a graph embedded without crossings as a straight line drawing in the plane, and we define the length of each edge in $G$ to be the Euclidean distance between its endpoints, then the \emph{dilation} of a pair of vertices $u,v$ in $G$ is the ratio between their distance in $G$ and their Euclidean distance. The dilation of $G$ is the maximum dilation of any two of its vertices.
A graph has dilation one (Figure~\ref{fig:dilation-free}) if and only if, between any two vertices, there exists a path connecting them that lies entirely along the line segment between them.

\begin{theorem}
\label{thm:quad-equivalence}
The following conditions are equivalent for a finite point set $P$:
\begin{enumerate}
\item $P$ has no empty quadrilateral.
\item $\quadgraph P$ is an independent set.
\item $P$ has only one triangulation.
\item $P$ is the set of vertices of a non-crossing embedded graph with dilation one.
\end{enumerate}
\end{theorem}

\begin{proof}
It is simplest to prove that the complements of these conditions are equivalent.

($\bar 1\Rightarrow\bar 2$): If $P$ has an empty quadrilateral, the two diagonals of this quadrilateral form an edge in the quadrilateral graph.

($\bar 1\Rightarrow\bar 3$): If $P$ has an empty quadrilateral, there exists a triangulation $T$ that includes the boundary edges of the quadrilateral and either one of its two diagonals (because any non-crossing set of line segments can be extended to a triangulation). But flipping the diagonal of this quadrilateral in $T$ produces a second triangulation; therefore, $P$ has more than one triangulation.

($\bar 1\Rightarrow\bar 4$): Suppose that $P$ has an empty quadrilateral. Then any dilation-one graph $G$ on $P$ would have to include line segments connecting opposite vertices of this quadrilateral; these line segments cross within the quadrilateral, contradicting the requirement that $G$ be embedded without crossings. Therefore, no such dilation-one graph exists.

($\bar 2\Rightarrow\bar 1$): If $\quadgraph P$ has an edge, its endpoints correspond to two crossing diagonals in $P$, and the four endpoints of these diagonals form an empty quadrilateral.

($\bar 3\Rightarrow\bar 1$): If $P$ has more than one triangulation, then perturb the point set by an affine transformation if necessary so that the Delaunay triangulation is unique; this perturbation does not affect the existence or nonexistence of empty quadrilaterals. After this perturbation, there exists a triangulation $T$ that is not Delaunay, because there are multiple triangulations but only one Delaunay triangulation. Then $T$ has a Delaunay flip, and the endpoints of the flipped edges form an empty quadrilateral.

($\bar 4\Rightarrow\bar 3$): Suppose $P$ is not the vertex set of a dilation-one graph, and let $T$ be any triangulation of $P$. Since no graph on $P$ has dilation one, this is in particular true of $T$; that is, there exists at least one pair of vertices in $P$ such that no path of $T$ lies along the line segment connecting these two vertices. Among pairs of vertices with this propertly, let $ab$ be the closest such pair. Then there can be no point of $P$ on line segment $ab$, for if such a point $c$ existed then $ac$ or $bc$ would be a closer pair with the same property. Thus, $ab$ must be a diagonal of $P$ that does not belong to $T$. There exists a triangulation $T'$ containing segment $ab$ (since any non-crossing set of line segments can be extended to a triangulation); $T$ and $T'$ are different triangulations since one of them contains $ab$ and the other does not.
\end{proof}

In previous work~\cite{Epp-97,DujEppSud-CGTA-07} we classified all point sets that can be vertex sets of dilation-one planar graphs. They are:
\begin{itemize}
\item $n$ collinear points (point sets with no empty triangle).
\item $n-1$ points on a line, and one point off the line.
\item $n-2$ points on a line, and two points on opposite sides of the lines, such that the $n-2$ points and the two points have disjoint convex hulls.
\item $n-2$ points on a line, and two points on opposite sides of the lines, such that the $n-2$ points and the two points have convex hulls that intersect at one of the $n-2$ points on the line.
\item A set of six points of a type shown in Figure~\ref{fig:dilation-free}.
\end{itemize}

Therefore, these are exactly the point sets that have no empty quadrilateral, and that have only one triangulation.

\section{Pentagons and partial cubes}

Any finite point set can be perturbed by a suitably chosen affine transformation in such a way that no four perturbed points are cocircular, without changing the orientation of any triple of points. In particular, such transformations preserve the set of triangulations of the point set as well as its flip graph. For this reason, we can assume when necessary throughout this section that the point set under consideration has no four cocircular points. With this assumption, we can orient each edge $ab$-$cd$ of $\quadgraph P$ from $ab$ to $cd$ whenever $\delaunay \{a,b,c,d\}$ has $cd$ as its diagonal. In the next sequence of lemmas we show that the assumption of no empty pentagon causes $\quadgraph P$ to have a particularly simple structure: it is a forest in which the edges in each tree are oriented towards a single vertex representing an edge of $\delaunay P$.

\begin{figure}[t]
\centering\includegraphics[width=3.5in]{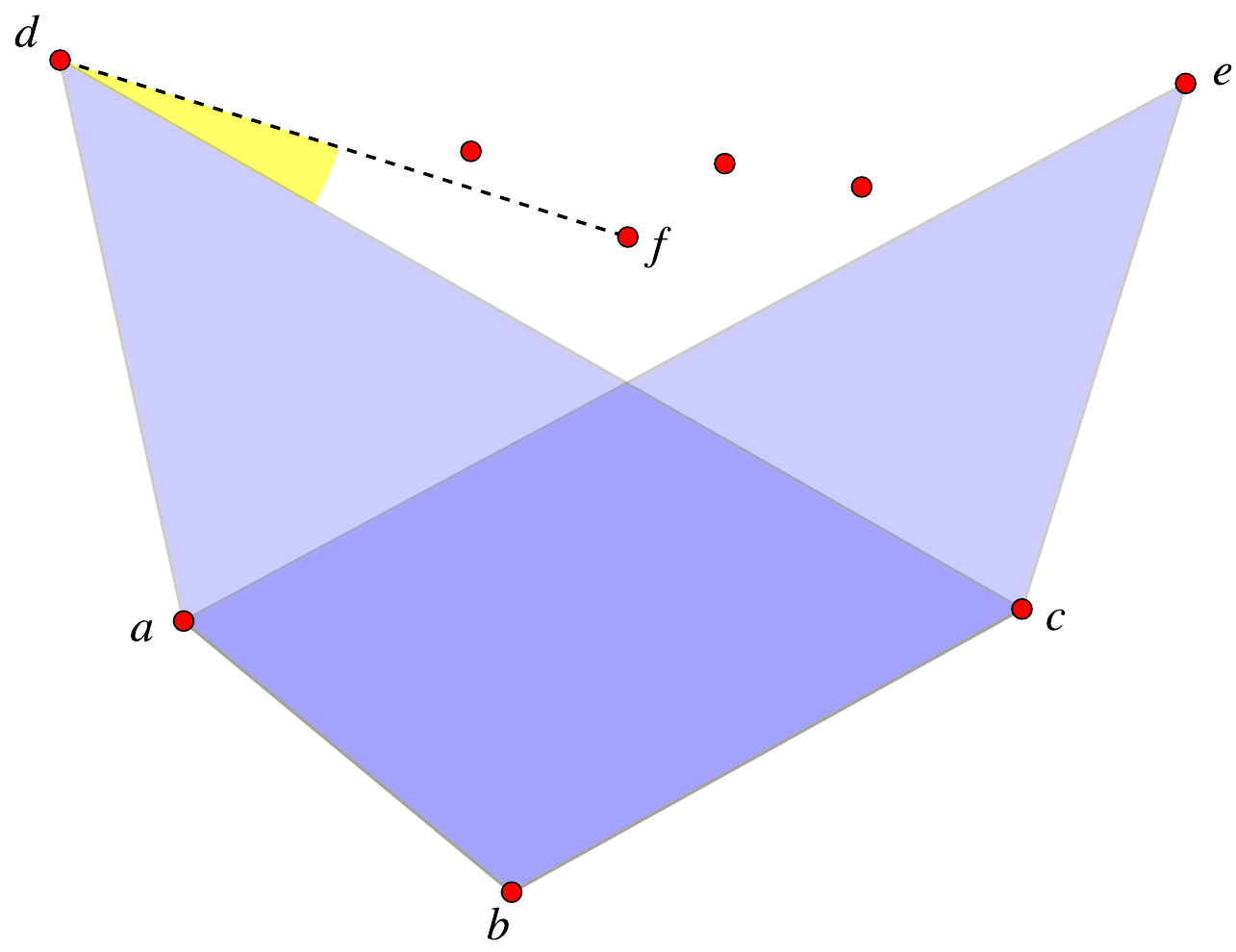}
\caption{Lemma~\ref{lem:pent-from-quads}: two empty quadrilaterals sharing two edges lead to an empty pentagon.}
\label{fig:pent-from-quads}
\end{figure}

\begin{lemma}
\label{lem:pent-from-quads}
If a finite point set $P$ contains two empty quadrilaterals that share two edges, then $P$ contains an empty pentagon.
\end{lemma}

\begin{proof}
The two shared edges must have a vertex in common, for otherwise it would not be possible to form two different quadrilaterals with those edges.
Therefore, let the two quadrilaterals have vertices $\{a,b,c,d\}$ and $\{a,b,c,e\}$. By the assumption that both quadrilaterals are empty, neither $d$ nor $e$ can be contained in the other quadrilateral. But the graph formed by the union of the edges of the two quadrilaterals is $K_{2,3}$, whose only planar drawings have one vertex inside the quadrilateral formed by the edges disjoint from it. Therefore, the six edges of the two quadrilaterals do not form a planar drawing, and there exists a pair of edges, one from each quadrilateral, that cross each other. Without loss of generality let these two crossing edges be $cd$ and $ae$, as shown in Figure~\ref{fig:pent-from-quads}.

Note that the set $S=(P\cap\hull\{c,d,e\})\setminus\{c,d\}$ is nonempty, because it contains $e$.
Among all points in $S$ choose $f$ to be one that minimizes angle $cdf$; if two or more points have the same minimum angle, choose $f$ to be the one closest to $d$.  Then $\hull\{a,b,c,d,f\}=\hull\{a,b,c,d\}\cup\hull\{c,d,f\}$. The quadrilateral $\hull\{a,b,c,d\}$ is empty by assumption, and the triangle $\hull\{c,d,f\}$ is empty because any point within it would contradict the choice of $f$ as minimizing angle $cdf$ or as being closest to $d$ with the given angle. Therefore $\hull\{a,b,c,d,f\}$ is empty, and forms an empty pentagon for $P$.
\end{proof}

\begin{lemma}
\label{lem:two-wedges}
Suppose that a finite point set $P$ contains no empty pentagon, that $a, b, c, d$ are the vertices of an empty quadrilateral in clockwise order, and that $\{a,c,e\}$ are the vertices of an empty triangle. Let $W_1$ be the closed wedge bounded by lines $ab$ and $ad$ that is (among the four wedges bounded by these two lines) the one opposite quadrilateral $abcd$, and symmetrically let $W_2$ be the closed wedge bounded by lines $bc$ and $cd$ that is (among the four wedges bounded by these two lines) the one opposite quadrilateral $abcd$. Then $e\in W_1\cup W_2$.
\end{lemma}

\begin{figure}[t]
\centering\includegraphics[width=5in]{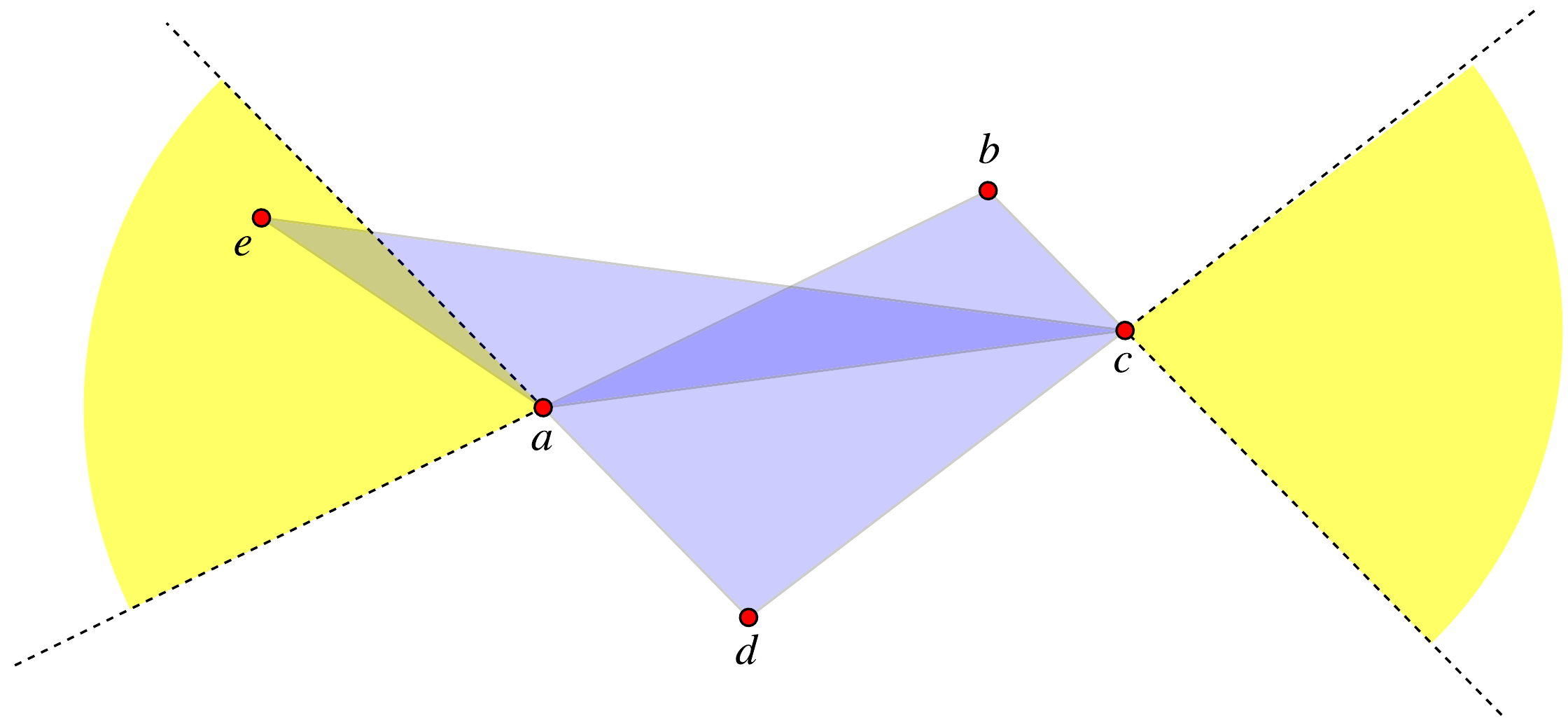}
\caption{Lemma~\ref{lem:two-wedges}: an empty quadrilateral restricts the location of the apex of an empty triangle to two wedges.}
\label{fig:two-wedges}
\end{figure}

\begin{proof}
For any other placement of $e$, either $abce$ or $aecd$ would be an empty quadrilateral and we could apply Lemma~\ref{lem:pent-from-quads} to find an empty pentagon, contradicting the assumption that $P$ contains no empty pentagon.
\end{proof}

An example of the lemma, showing the two wedges in which $e$ must be contained, is illustrated in Figure~\ref{fig:two-wedges}.

\begin{figure}[t]
\centering\includegraphics[width=5in]{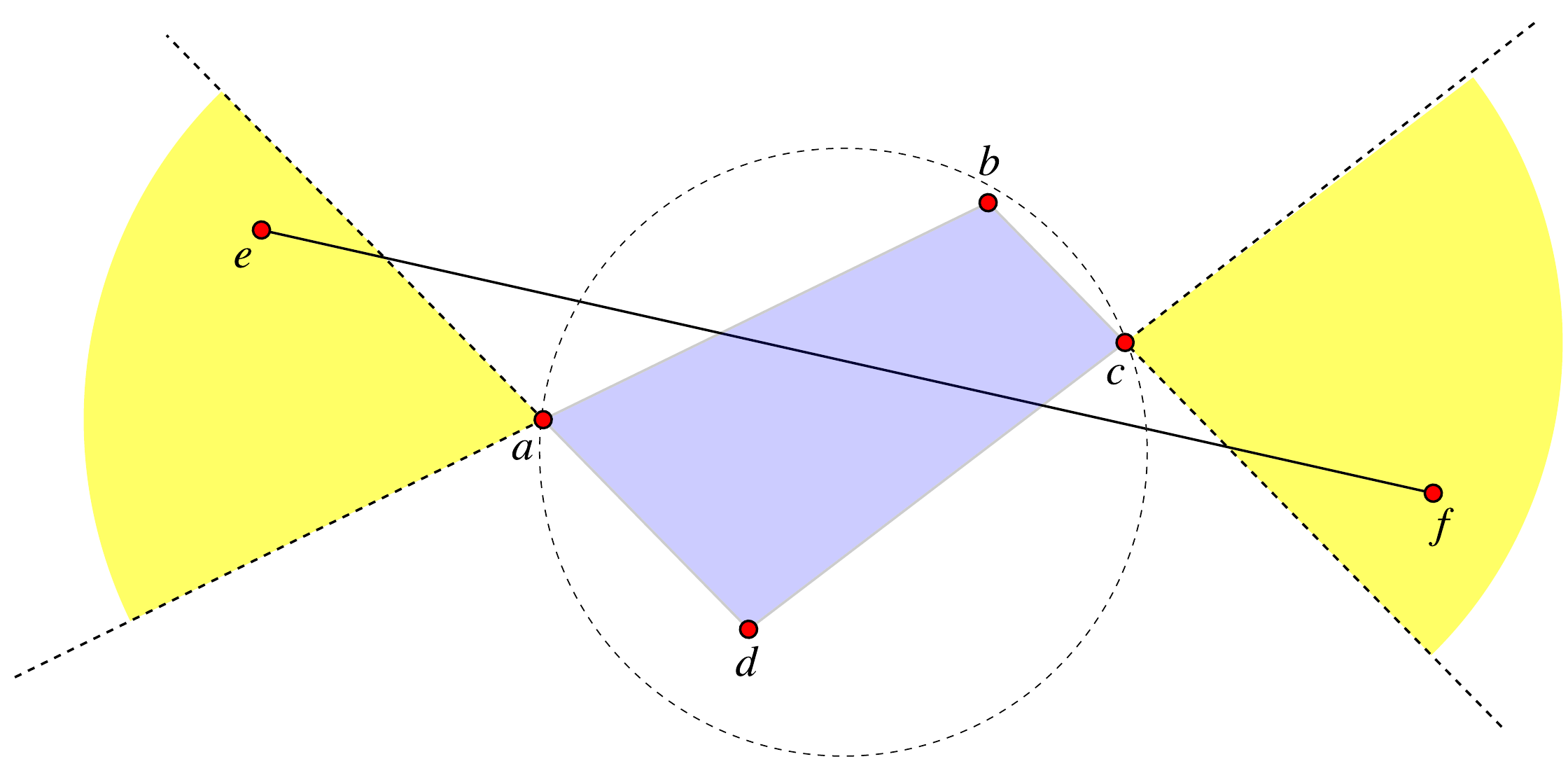}
\caption{Lemma~\ref{lem:DT-uniqueness}: in a set with no empty pentagons, if $bd$ is a Delaunay flip for $ac$, then there can be no other Delaunay flip $ef$.}
\label{fig:DT-uniqueness}
\end{figure}

\begin{lemma}
\label{lem:DT-uniqueness}
Let finite point set $P$ contain no empty pentagon and have no four cocircular points, and let $ac$ be a diagonal of $P$ that does not belong to $\delaunay P$. Then $ac$ has a unique Delaunay flip $bd$.
\end{lemma}

\begin{proof}
Let $bd$ be any Delaunay flip for $ac$.
Diagonal $bd$ belongs to $\delaunay\{a,b,c,d\}$ if and only if there exists a circle $O$ having $a$ and $c$ on its boundary and $b$ and $d$ in its interior, if and only if the sum of angles $abc$ and $adc$ is at least $\pi$. If $ac-ef$ is any edge in the quadrilateral graph of $P$, with $ef\ne bd$,
then by Lemma~\ref{lem:two-wedges} we must have $e$ and $f$ belonging to the two wedges $W_1$ and $W_2$. Necessarily one of $e$ and $f$ must belong to each wedge, for otherwise $ef$ would not cross $ac$ contradicting the assumption that $ac-ef$ belongs to the quadrilateral graph.

Wedge $W_1$ is bounded by a ray with apex $a$ along line $ab$ and on the opposite side of $a$ from $b$; this line crosses $O$ twice, once at $a$ and once on the other side of $b$ from $a$. Therefore, the boundary ray of $W_1$ is entirely outside $O$. By a symmetric argument, the other three boundary rays of $W_1$ and $W_2$ are also entirely outside $O$. Therefore $W_1$ and $W_2$ themselves are exterior to $O$ (Figure~\ref{fig:DT-uniqueness}). Since $O$ is an empty circle for $ac$ among the set of four sites $\{a,c,e,f\}$, $ac$ must be an edge of $\delaunay\{a,c,e,f\}$ and $ef$ cannot be a Delaunay flip for $ac$.
\end{proof}

\begin{lemma}
\label{lem:QG-is-tree}
If a finite point set $P$ contains no empty pentagon, $\quadgraph P$ is a forest.
\end{lemma}

\begin{proof}
We perturb $P$ by an affine transformation if necessary so that no four points are cocircular; this perturbation does not change $\quadgraph P$.
For each edge $ac$-$bd$ of $\quadgraph P$, either $bd$ is a Delaunay flip of $ac$ or vice versa.
We orient each edge from $ac$ to $bd$ if $bd$ is the Delaunay flip of $ac$, and the other orientation otherwise. Repeated Delaunay flipping must always lead to the Delaunay triangulation~\cite{Law-72,Sib-CJ-73}, so this orientation is acyclic and (by Lemma~\ref{lem:DT-uniqueness}) has at most one outgoing edge per vertex. Therefore each connected component of $\quadgraph P$ is a tree and $\quadgraph P$ is a forest.
\end{proof}

\begin{lemma}
\label{lem:one-per}
If finite point set $P$ contains no empty pentagon, and $T$ is any triangulation of $P$, then $T$ contains exactly one edge from each tree in $\quadgraph P$.
\end{lemma}

\begin{proof}
We perturb $P$ by an affine transformation if necessary so that no four points are cocircular; this perturbation does not affect the statement of the lemma.
According to the orientation from the proof of Lemma~\ref{lem:QG-is-tree}, each tree in $\quadgraph P$ has a unique Delaunay edge (the edge at the root of the tree), so the lemma holds when $T$ is the Delaunay triangulation.
For arbitrary $T$ the result follows by induction on the number of flips needed to reach the Delaunay triangulation: each Delaunay flip reduces this number by one and preserves the number of edges per tree in $\quadgraph P$.
\end{proof}

\begin{figure}[t]
\centering\includegraphics[width=3.5in]{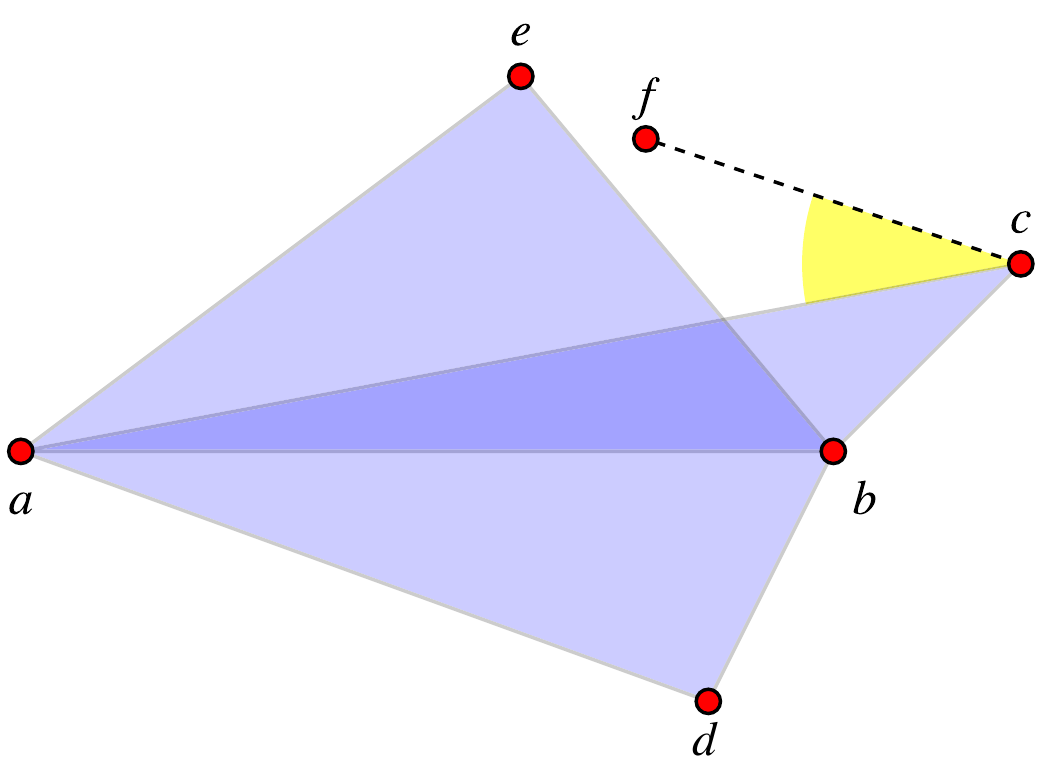}
\caption{Lemma~\ref{lem:dflip-endpt}: if $abc$ is a non-Delaunay empty triangle, then one of its vertices must be an endpoint of the Delaunay flip for the opposite edge.}
\label{fig:dflip-endpt}
\end{figure}

\begin{lemma}
\label{lem:dflip-endpt}
Let finite point set $P$ have no empty pentagon and no four cocircular points, and let $abc$ be an empty triangle that is not part of $\delaunay P$. Then at least one vertex of $abc$ is an endpoint of the Delaunay flip of the opposite edge.
\end{lemma}

\begin{proof}
If $c$ is an endpoint for the Delaunay flip for $ab$, the result holds; otherwise, suppose that the Delaunay flip for $ab$ is $de$. Swap the names of points $a$, $b$ or of points $d$, $e$ if necessary so that $ac$ crosses the boundary of quadrilateral $adbe$ on edge $be$ (Figure~\ref{fig:dflip-endpt}). Let
$S$ be the set $(P\cap\hull\{a,c,e\})\setminus \{a,c\}$; $S$ is non-empty, because it contains $e$.
Let $f$ be the point of $S$ that minimizes angle $acf$; if multiple points have the same minimizing angle, choose $f$ to be the one closest to $c$. Then $abcf$ is an empty quadrilateral, for any point inside it would form a sharper angle than $f$ or be closer to $c$ with the same angle, contradicting the choice of $f$.

Angle $afc$ is at least equal to angle $aec$, which exceeds angle $aeb$. Moreover, angle $abc$ exceeds angle $adb$, for otherwise $c$ would not belong to the wedge for empty quadrilateral $adbe$ described in Lemma~\ref{lem:two-wedges}. Therefore, $afc+abc > aeb+adb > \pi$ (where the second inequality $aeb+adb > \pi$ characterizes the  fact that $de$ is a Delaunay flip for $ab$); since $afc+abc>\pi$, $bf$ is a Delaunay flip for $ac$. Therefore, $b$ is an endpoint for the Delaunay flip of the opposite edge $ac$.
\end{proof}

Define the \emph{flip number} $\# ab$ of a diagonal $ab$ in a point set $P$ with no four cocircular points to be the minimum number of flips needed to reach a triangulation containing $ab$ from $\delaunay P$. If $ab$ is a Delaunay triangulation edge, define $\# ab = 0$.

\begin{lemma}
\label{lem:max-flip-is-delaunay}
Let $P$ be a finite point set with no empty pentagon and no four cocircular points.
Suppose that $abc$ and $acd$ are empty triangles, intersecting only on non-Delaunay edge $ac$, such that $\# ac \ge \max\{\# ab, \# bc, \# cd, \# ad\}$. Then $bd$ is the Delaunay flip of $ac$.
\end{lemma}

\begin{proof}
If $ab$ and $bc$ are both Delaunay edges, then any sequence of Delaunay flips that produces the Delaunay triangulation from some triangulation containing $ac$ must eventually flip $ac$, without ever flipping $ab$ and $bc$; therefore $b$ is an endpoint of the Delaunay flip for $ac$.
If on the other hand $ab$ is not a Delaunay edge, then $b$ must be an endpoint of the Delaunay flip for $ac$, for otherwise by Lemma~\ref{lem:dflip-endpt} $ac$ would be an edge of the convex hull of the four vertices forming the Delaunay flip for $ab$ or $bc$, so any flip sequence producing that edge from the Delaunay triangulation would have to contain $ac$ before its final step, contradicting the assumption that $\# ac \ge \max\{\# ab,\# bc\}$. Thus regardless of which edges are Delaunay, $b$ is an endpoint of the Delaunay flip for $ac$.
By a symmetric argument $d$ is also an endpoint of the Delaunay flip; therefore the flip is $bd$.
\end{proof}

\begin{theorem}
\label{thm:pent-equivalence}
The following three conditions are equivalent for a finite point set $P$:
\begin{enumerate}
\item $P$ has no empty pentagon.
\item $\quadgraph P$ is a forest.
\item $\flipgraph P$ is a partial cube.
\end{enumerate}
\end{theorem}

\begin{proof}
($2\Rightarrow 1$): If $\quadgraph P$ is a forest, $P$ must have no empty pentagon, because the diagonals of an empty pentagon would form a cycle of five vertices in $\quadgraph P$, but a forest has no cycles.

($1\Rightarrow 2$): Lemma~\ref{lem:QG-is-tree}.

($3\Rightarrow 1$): If $\flipgraph P$ is a partial cube, $P$ must have no empty pentagon, because the diagonals of an empty pentagon can be flipped one for another to form a cycle of five vertices in $\flipgraph P$, but every partial cube is bipartite.

($1\Rightarrow 3$): In the remainder of the proof we show that, if $P$ has no empty pentagon, $\flipgraph P$ is a partial cube. Let $\Pi$ be the partial cube formed as the Cartesian product of the trees in the forest $\quadgraph P$.
A vertex in $\Pi$ can be identified by selecting one vertex for each tree in $\quadgraph P$.  By Lemma~\ref{lem:one-per}, any triangulation $T$ has one diagonal corresponding to a vertex in each tree of $\quadgraph P$, so $T$ corresponds to a vertex in $\Pi$. We show below that this identification of triangulations with vertices in $\Pi$ is isometric, by showing both upper and lower bounds relating flip distance to distance in $\Pi$.

We begin by showing that flip distance is at least equal to distance in $\Pi$.
If two triangulations $T_1$ and $T_2$ differ by a flip, then the vertices in $\quadgraph P$ representing their sets of diagonals differ only in a single tree of $\quadgraph P$, and their images of $T_1$ and $T_2$ in $\Pi$ are adjacent.
More generally, let $T_1$ and $T_2$ be any two triangulations of $P$, at flip distance $\delta$. In any sequence of $\delta$ flips changing $T_1$ to $T_2$, each flip corresponds in $\Pi$ to an adjacency between vertices, so (by the triangle inequality for distances in $\Pi$) the distance between the images of $T_1$ and $T_2$ in $\Pi$ is at most $\delta$.

To complete the proof, we show that flip distance is at most equal to distance in $\Pi$, by showing that any two triangulations can be connected by a short sequence of flips. Suppose $T_1$ and $T_2$ have images in $\Pi$ that are $\delta$ apart. We perturb the input by an affine transformation if necessary so that no four points are cocircular, allowing us to define Delaunay flips and flip numbers; let $ab$ be the edge in the symmetric difference of $T_1$ and $T_2$ maximizing $\# ab$.  The two triangles based on this edge (in whichever of $T_1$ or $T_2$ contains the edge) meet the preconditions of Lemma~\ref{lem:max-flip-is-delaunay}, so there exists a Delaunay flip of $ab$ in the triangulation $T_i$ containing $ab$. This Delaunay flip must change $T_i$ to a triangulation in which the representative vertex in $\quadgraph P$ is closer than $ab$ to the representative vertex in the other triangulation, so the flipped triangulation has distance $\delta-1$ from the other triangulation. By repeating this choice of flipping the edge with the maximum flip number, we can find a sequence of $\delta$ flips that transforms $T_1$ to $T_2$, so the flip distance is at most equal to $\delta$.

Since the flip distance and the distance in $\Pi$ are both upper bounded by the other, they are equal. Since we have embedded $\flipgraph P$ isometrically in a partial cube $\Pi$, $\flipgraph P$ is itself a partial cube.
\end{proof}

For a set $P$ of $n$ points with no empty pentagon, $\quadgraph P$ has a number of edges less than its number of vertices, as it is a forest. But there are $O(n^2)$ vertices in $\quadgraph P$, because each corresponds to a pair of points in $P$. Thus, in such a point set, there can be at most $O(n^2)$ empty quadrilaterals. Placing $n/2$ points on each of two parallel lines forms a set of $n$ points with no empty pentagon and $(n/2-1)^2$ empty quadrilaterals, showing that this quadratic bound is tight.

\section{Examples}

We describe briefly some examples of finite point sets without empty pentagons, showing that they exhibit great variety despite the need for some collinearities. We do not have a complete classification of such sets, but they appear to allow significantly greater complexity than the point sets without empty quadrilaterals classified in earlier work and described earlier in this paper.

\begin{figure}[t]
\centering\includegraphics[width=6in]{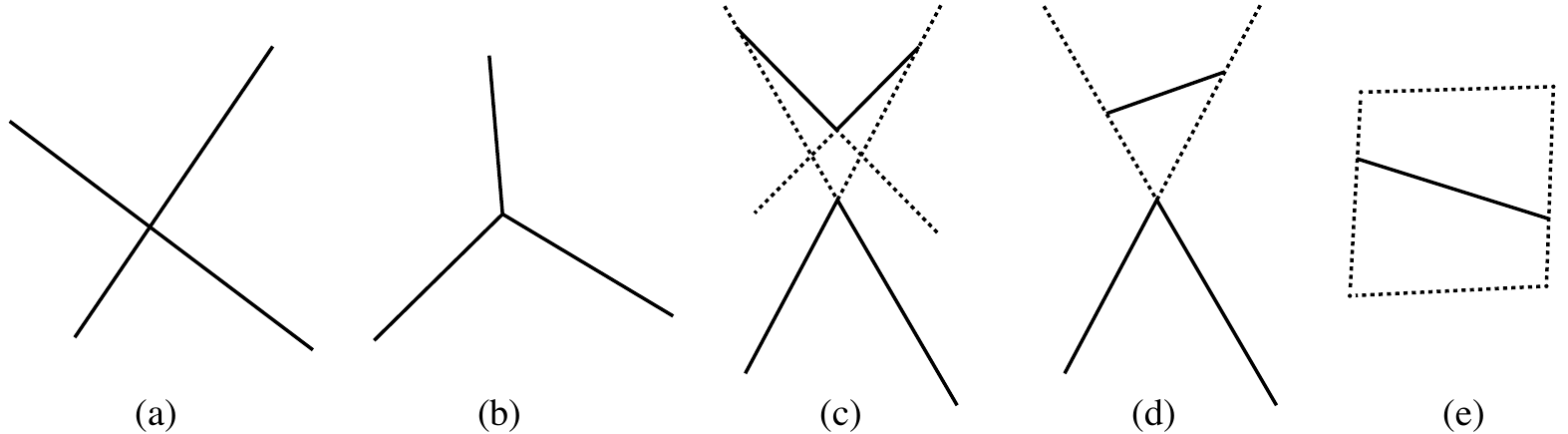}
\caption{Continuous point sets with no pentagons: (a) two lines, (b) three rays, (c) two opposite wedges, (d) a wedge and a line segment, and (e) a quadrilateral and a line segment.}
\label{fig:no-pent}
\end{figure}

\begin{itemize}
\item Any point set with no empty quadrilateral clearly also has no empty pentagon. Thus, all the types of point sets with no empty quadrilateral, as described in Section~\ref{sec:no-empty-quad}, also form examples of point sets with no empty pentagon.

\item If $P$ is a point set with no empty pentagons, and $C$ is any convex set in the plane, then $P\cap C$ again has no empty pentagons. In addition, any projective transformation of $P$ that preserves the orientation of all its triangles results in a point set with no empty pentagons.

\item The union of two lines has no convex pentagons at all, let alone empty convex pentagons (Figure~\ref{fig:no-pent}, far left). Therefore, one can form finite sets of points with no convex pentagons by taking any finite subset of the union of two lines.

\item Similarly, if three rays share a common apex and the three wedges bounded by these rays each form angles less than $\pi$ (Figure~\ref{fig:no-pent}, near left), the points from these three rays do not form any convex pentagons, and one can form finite point sets with no convex pentagon by taking any  finite subset of these three rays.

\item Let $W_1$ and $W_2$ be two wedges, with $\bar W_1$ and $\bar W_2$ the complementary wedges bounded by the other two rays on the same lines. Suppose further that the apex of $W_1$ is contained in $\bar W_2$ and the apex of $W_2$ is contained in $\bar W_1$. Let $V_1$ be a subset of the boundary of $W_1$ that is entirely contained in $\bar W_2$, and let $V_2$ be a subset of the boundary of $W_2$ that is entirely contained in $\bar W_1$ (Figure~\ref{fig:no-pent}, center). Then $V_1\cup V_2$, and any of its finite subsets, do not contain the vertices of any convex pentagon.

\item Let $V$ be the boundary of a wedge $W$, and $S$ be a line segment contained in the complementary wedge $\bar W$, such that the line containing $S$ is disjoint from $W$ (Figure~\ref{fig:no-pent}, near right). Then again $V\cup S$ and its finite subsets do not form any convex pentagons.

\item The vertices of a convex quadrilateral, and any subset of a line segment that connects two opposite sides of the quadrilateral, do not form any convex pentagon (Figure~\ref{fig:no-pent}, far right).

\item Any lattice in the plane (for instance the lattice of points with integer coordinates) has no empty pentagons~\cite{Ark-BAMS-80,KloAleLar-AMM-91,LarGil-MM-93}. This follows immediately from the observation that there are only four possible combinations of the parity of the $x$-coordinate and the parity of the $y$-coordinate of a grid point. Therefore, among any five grid points, some two have the same parity for both coordinates, and their midpoint is also a grid point, causing the pentagon formed by the five points to be non-empty. The stronger assertion that any lattice pentagon $abcde$ has another lattice point within the small pentagon $S(a,b,c,d,e)$ formed by its inner diagonals  (Figure~\ref{fig:gridpent}) has supposedly appeared as a problem in the Russian mathematics olympiad, and can be solved by a simple case analysis:

\begin{itemize}
\item Suppose we have five points $a,b,c,d,e$ forming (in clockwise order) a convex pentagon in a grid, with $m$ the midpoint of some two points with the same parity. If $m$ lies within $S(a,b,c,d,e)$, as shown in the left and center of Figure~\ref{fig:gridpent}, we are done.
\item If $m$ lies on one of the internal diagonals of pentagon $abcde$, but does not belong to the inner pentagon $S(a,b,c,d,e)$, we may assume without loss of generality that $m$ is the midpoint of segment $ac$. Then one of the two pentagons $abmde$ or $mbcde$ is convex (depending on the relative position of the inner pentagon and $m$ on segment $ac$) and has an inner pentagon that is a subset of  $S(a,b,c,d,e)$. By induction on the number of grid points contained in the given pentagon, this new smaller pentagon has a grid point in its inner pentagon, which must therefore also belong to $S(a,b,c,d,e)$.
\item In the remaining case, $m$ is a midpoint of one of the outer edges of the pentagon, as shown in the center and right of Figure~\ref{fig:gridpent}; without loss of generality $m$ lies on edge $ab$. Now, $S(m,b,c,d,e)$ is not in general a subset of $S(a,b,c,d,e)$, but it is a subset of 
$S(a,b,c,d,e)\cup T$ where $T$ is the triangle formed by lines $ac$, $bd$, and $be$.  
As before, by induction, $S(m,b,c,d,e)$ contains a lattice point $f$.  If $f$ is 
in $S(a,b,c,d,e)$, we are done.  If $f$ is not in $S(a,b,c,d,e)$, it is in $T$, and 
$afcde$ is a convex pentagon such that $S(a,f,c,d,e)$ is a subset of $S(a,b,c,d,e)$;  
by induction a third time, $S(a,f,c,d,e)$ contains a lattice point, which is 
also a lattice point in $S(a,b,c,d,e)$.
\end{itemize}

\begin{figure}[t]
\centering\includegraphics[width=5in]{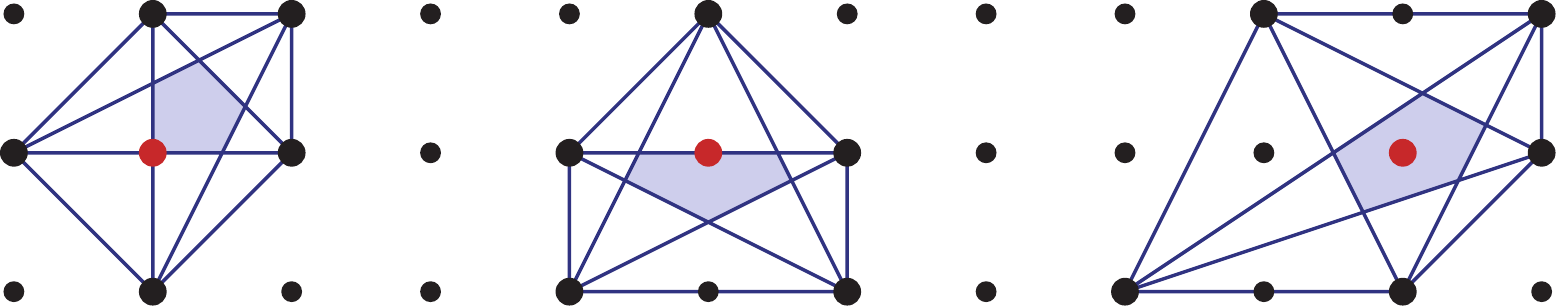}
\caption{Any lattice pentagon has another lattice point in it, and more specifically in the smaller closed pentagon formed by its inner diagonals.}
\label{fig:gridpent}
\end{figure}

Thus, intersecting the lattice with a bounded convex set results in a finite set $P$ without empty pentagons. An example of a flip graph for a point set formed from a $3\times 3$ square subset of a lattice can be seen in Figure~\ref{fig:3x3flipgroups}.

\item One may also intersect the lattice with a bounded convex set $K$ and then remove any subset of points that lie on the boundary of the convex hull of the intersection, without changing the property of having no empty pentagons. The resulting reduced set $P$ has no empty pentagon. To see this, suppose that $E$ is any empty convex polygon in $P$. We consider the following cases:
\begin{itemize}
\item If none of the removed boundary points lies on the boundary of $E$, then the grid points in $E$ are the same as those in the intersection of the grid with $K$, which has no empty convex pentagon as above. Therefore, $E$ has at most four sides.
\item If $E$ has two consecutive sides $ab$ and $bc$ from which boundary points have been removed, let $p$ be the removed point closest to $b$ on side $ab$ and let $q$ be the removed point closest to $c$ on side $ac$, and consider the grid point $r=p+q-b$ (the fourth point of a parallelogram with vertices $p$, $b$, and $q$). Because distance $ab$ is an integer multiple of distance $pb$, $a$ is at least twice as far from $b$ as $p$ is, and similarly $c$ is at least twice as far from $b$ as $q$ is. Therefore, $r$ must lie within the convex hull of $abc$, so the only way for $E$ to be empty is for $abc$ to be a triangle with $r$ on its boundary edge $ac$.
, we may find an affine transformation of the plane such that that the lattice from which $P$ is formed is the integer lattice, with these two consecutive sides of $E$ on the coordinate axes of the plane and $E$ entirely within the positive quadrant. That is, the two sides of $E$ are the line segments from the origin to the points $(x,0)$ and $(0,y)$ for some integers $x$ and $y$, both greater than one. But then $E$ must contain the point $(1,1)$; for $E$ to be an empty polygon with respect to $P$, the point $(1,1)$ must be one of the removed points, $x=y=2$, and $E$ must be a triangle.
\item In the remaining case $E$ has a single side $bc$ from which a boundary point has been removed, let $ab$ be the other side of $E$ with vertex $b$, let $p$ be the removed point closest to $b$ on segment $bc$, and let $q=a+p-b$ be the fourth point of a parallelogram with vertices $a$, $b$, and $p$. This parallelogram cannot contain any grid points other than its vertices, because if it did then by point-reflection symmetry the half-parallelogram $abp$ would contain a grid point, which would lie in the interior of $E$ and therefore could not be one of the removed points; this contradicts the assumption that $E$ is empty. Thus, $abpq$ is a period parallelogram of the integer lattice, and we may perform an affine transformation (without changing the emptiness of $E$) to bring $a$ to the point $(0,1)$, $b$ to $(0,0)$, $p$ to $(1,0)$, and $q$ to $(1,1)$. Then $E$ is entirely contained within the positive quadrant, and all vertices of $E$ must lie on the two lines $y=0$ and $y=1$, because for any other point $r$ in the positive quadrant the convex hull of $abcr$ would contain $q$ in its interior. But any convex polygon whose vertices lie on two lines can only be a triangle or a quadrilateral.
\end{itemize}
Therefore, as claimed, the sets formed in this way also have the empty pentagon property.

\begin{figure}[t]
\centering\includegraphics[width=5.5in]{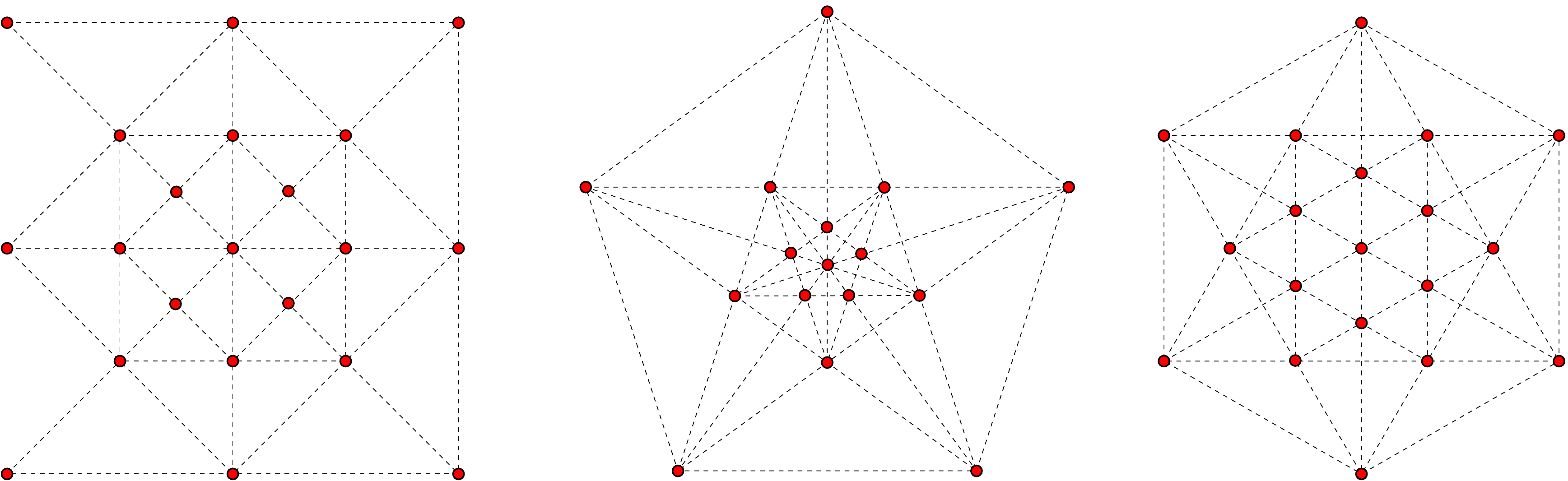}
\caption{Point sets with no empty pentagon derived from nested regular polygons.}
\label{fig:nested}
\end{figure}

\item The point sets in Figure~\ref{fig:nested}, derived from nested regular polygons, do not have any empty pentagons. In each case an additional level of nesting would lead to an empty pentagon.

\item Let $P$ be a finite subset of the union of two lines, and let $Q$ be a finite subset of a single line,
such that any empty triangle of $P$ contains a point from $Q$. 
Then $P\cup Q$ contains no empty pentagons, for any pentagon must have three noncollinear vertices from $P$ and some point from $Q$ would cause it to be nonempty (Figure~\ref{fig:trihit}).
Similar constructions are possible in which $P$ is a subset of one of the other pentagon-free configurations in Figure~\ref{fig:no-pent}, and $Q$ is an appropriate subset of a line segment separating the two components of the configuration.

\item K\'ara et al.~\cite{KarPorWoo-DCG-05} observe that the infinite set of points with Cartesian coordinates $(12i, 6)$, $(3i, 3)$, $(4i, 2)$, or $(6i, 0)$ for integer values of~$i$ form a set with no empty pentagon; the same is true for the finite sets formed by intersecting this infinite set with any bounded convex set.
\end{itemize}

\begin{figure}[t]
\centering\includegraphics[width=3.75in]{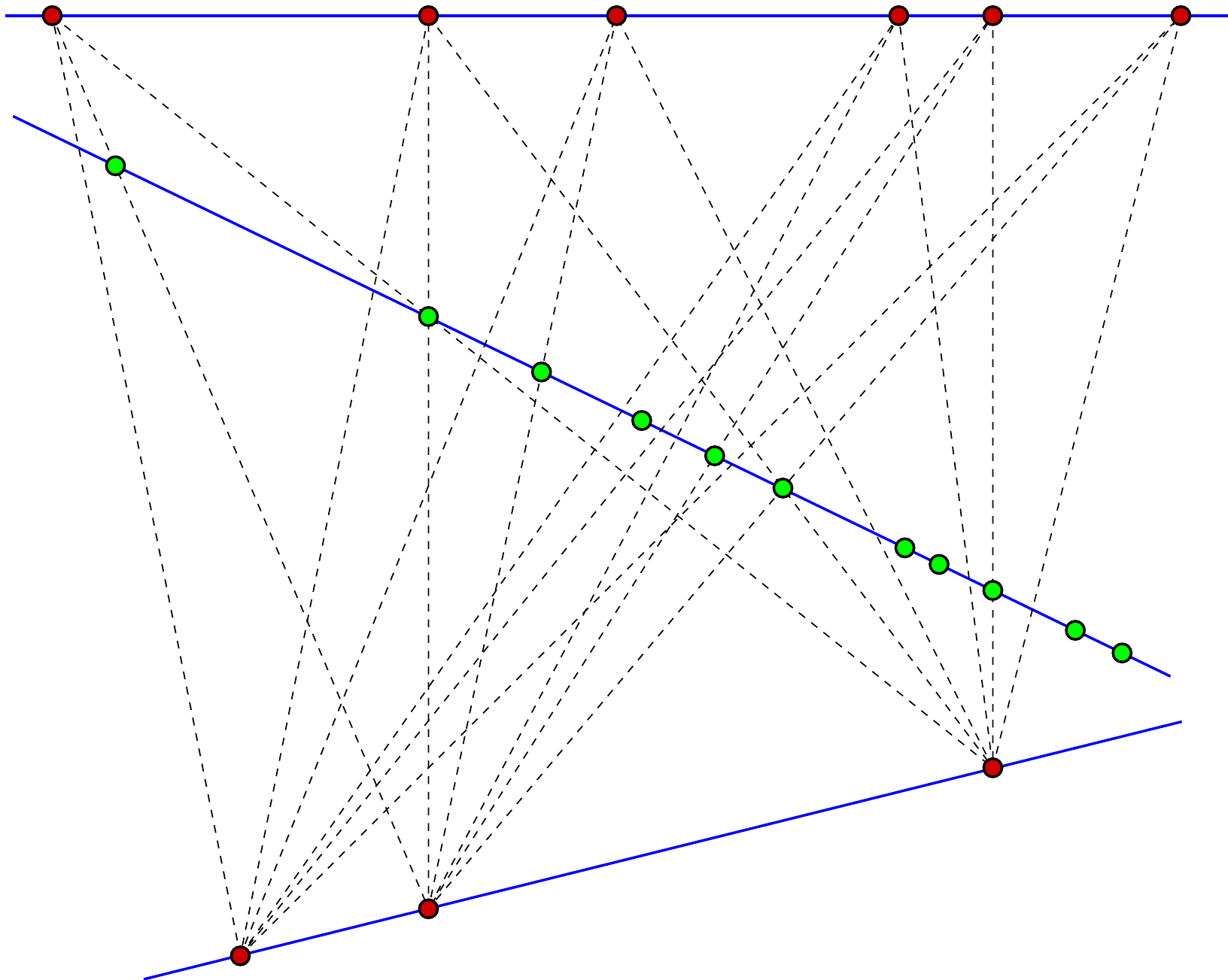}
\caption{A finite subset $P$ of the union of two lines (the dark circles) and a collinear point set $Q$ (light circles) that includes a point in every empty triangle of $P$.}
\label{fig:trihit}
\end{figure}

\section{Finding pentagons and flips}

Our classification of point sets with no empty convex quadrilateral allows us to determine whether a point set has an empty quadrilateral, in linear time. For larger constant values of $k$, an algorithm of Eppstein \emph{et al.}~\cite{EppOveRot-DCG-92} allows the existence of an empty convex $k$-gon to be determined in time $O(n^3)$. We now describe a simple technique that solves the problem of testing for the existence of an empty pentagon more quickly, in $O(n^2)$ time.

\begin{lemma}
\label{lem:test-bottom}
Let $q$ be the unique point with minimum $y$-coordinate in a point set $P$, and suppose that the remaining points are given in sorted order according to the slopes of the lines they form with $q$. Then we can find an empty pentagon in $P$ with $q$ as one of its vertices, if such a pentagon exists, in linear time.
\end{lemma}

\begin{figure}[t]
\centering\includegraphics[width=5.5in]{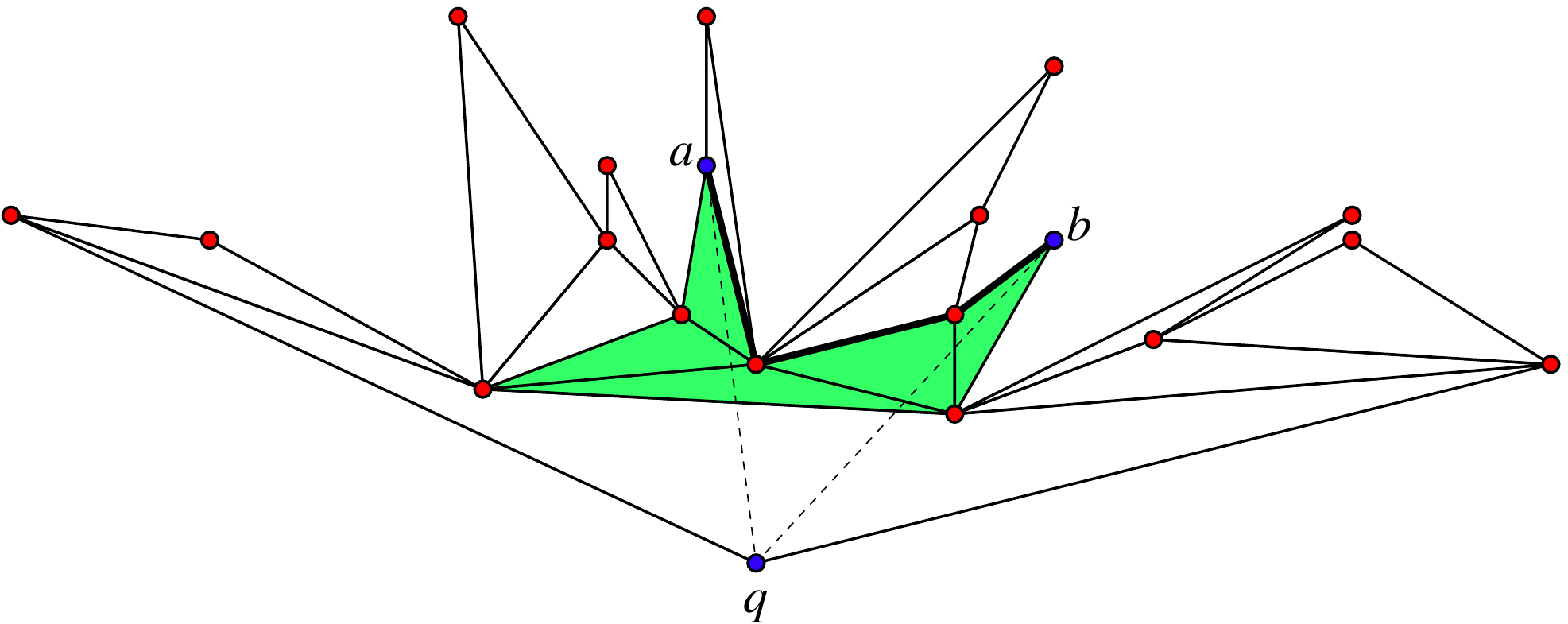}
\caption{A set of points, connected in radial order as viewed from its bottom vertex $q$ to form a polygon, and partitioned into triangles by a maximal set of diagonals disjoint from $q$. No two adjacent triangles form a convex quadrilateral, from which it follows that any two vertices $a$ and $b$ can be connected by a concave chain of diagonals. The shaded region corresponds to a path connecting $a$ to $b$ in the dual graph of the partition.}
\label{fig:starnopent}
\end{figure}

\begin{proof}
If two points in $P$ form the same slope with $q$, remove all but the point nearest $q$; this does not change the existence of an empty pentagon with $q$ as vertex, because none of the removed points can be in an empty polygon with $q$.
Connect the remaining points by a polygonal chain in the sorted order, and connect the ends of this chain to $q$; this forms a simple polygon each point of which is visible from $q$. As long as this polygon contains a concave vertex other than $q$, simplify the polygon by removing such a vertex and adding a diagonal connecting its two neighbors; this simplification process produces a partition of the initial polygon into a collection of triangles, disjoint from $q$, and a \emph{pseudotriangle} (a simple polygon with three non-reflex vertices) of which $q$ is a vertex (Figure~\ref{fig:starnopent}). The planar dual graph of this partition is a tree, with the pseudotriangle containing $q$ as root, and each triangle of this partition is visible from $q$ through the triangle edge connecting it to its parent in this dual tree.

We assert that $P$ has an empty pentagon, with $q$ as vertex, if and only if some two adjacent triangles in this partition (neither having $q$ as vertex) form a convex quadrilateral. In one direction, it is clear that any such quadrilateral forms with $q$ a convex pentagon. In the other direction, suppose that there is no convex quadrilateral of this type, and consider any four points $a$, $b$, $c$, $d$ in $P\setminus\{q\}$. As the diagonals added in constructing our partition form an outerplanar graph, some two of these points, say $a$ and $b$, must not be connected by a diagonal of the partition, because outerplanar graphs cannot contain a complete graph $K_4$ on four vertices. Let $A$ be the region containing $a$ that is closest to $q$ in the partition, and similarly let $B$ be the region containing $B$ that is closest to $q$. Let $U$ denote the path in the tree dual to the partition connecting $A$ to $B$, corresponding to a collection of regions in our partition (shown shaded in the figure), and connect $a$ to $b$ by a path $u$ of diagonals along the part of the boundary of $U$ farthest from $q$.
Then, by the condition that no two adjacent triangles in the partition be convex, $u$ follows a concave angle at each of its vertices, and $u$ together with edges $qa$ and $qb$ forms a pseudotriangle. As $\hull\{q,a,b\}$ is nonempty, $a$, $b$, $c$, $d$, and $q$ cannot form part of an empty pentagon. Therefore, there can be no empty pentagon with $q$ as boundary.

The partition into triangles, and the test for the existence of two adjacent triangles forming a convex quadrilateral, are easily performed in linear time.
\end{proof}

To perform this computation efficiently, we need for each point $q$ the radial sorted order of the points above it. The algorithm below finds these sorted orders efficiently by applying projective duality to transform the problem into one of constructing a line arrangement~\cite{ChaGuiLee-BIT-85}; it is by now a standard component in the computational geometry repertoire, but we include it for completeness.

\begin{lemma}
\label{lem:radial-sort}
Suppose we are given a set of points, no two of which have the same $y$-coordinate. Then we may determine, for each point $q$ in the set, the radial sorted order around $q$ of the points with greater $y$-coordinate, in a total time of $O(n^2)$.
\end{lemma}

\begin{proof}
We assume without loss of generality (by perturbing the points if necessary) that no two input points determine a vertical line. If the perturbation consists of a small enough rotation of the plane, this will not change the radial sorted ordering; here ``small enough'' means that the $y$-coordinate ordering does not change. To find a suitable rotation angle, find a line $\ell$ determined by some two input points that is neither horizontal nor vertical, but that forms an angle that is as sharp as possible with one of the coordinate axes; this may easily be done in $O(n^2)$ time. Then, a rotation of the input set by half the angle formed by $\ell$ with a coordinate axis will preserve the $y$-ordering of the points and eliminate all vertical lines determined by pairs of points.

We consider two planes, which we call the primal and the dual.
The primal plane is given by coordinates $x,y$, and contains (non-vertical) lines defined by equations of the form $y=mx+b$, for pairs of numbers $m,b$. Analogously, the dual plane is given by coordinates $X,Y$, and contains (non-vertical) lines defined by equations of the form $Y=MX+B$, for pairs of numbers $M,B$. To each of our input points $(x,y)$ we associate the dual line determined by the numbers $M=-x,B=y$, and to each line $y=mx+b$ in the primal plane we associate the dual point $X=m,Y=b$. Then, with this association, $y=mx+b$ if and only if $Y=MX+B$; that is, an input point lies on some primal line if and only if the line dual to the input point contains the point dual to the primal line.
Thus, the dual points where any two dual lines cross correspond to the primal lines determined by two input points, and the sorted order of these dual points by their $X$-coordinates (that is, the order in which they appear along the dual lines) is the same as the slope ordering of the lines determined by pairs of points in the primal plane.

Thus, by forming the set of lines in the dual plane corresponding to the input points in the primal plane, and applying an algorithm for constructing line arrangements~\cite{ChaGuiLee-BIT-85}, we may find in total time $O(n^2)$, for each point $q$, the sorted order of all the other points by the slopes of the lines they form with $q$. We may then remove from this sorted order the points below $q$, and move the subsequence of lines with negative slope to the start of the order, to form the radial ordering of the points above $q$.
\end{proof}

\begin{theorem}
\label{thm:test-pent}
We may test whether any set of $n$ points contains an empty pentagon, in time $O(n^2)$.
\end{theorem}

\begin{proof}
We rotate the point set if necessary, as in the proof of Lemma~\ref{lem:radial-sort}, so that no two points have the same $y$-coordinate.
We then apply Lemma~\ref{lem:test-bottom} to each point $q$ in the point set, and to the set of points having larger $y$-coordinates than each $q$. Sorting all of the input points radially around each point $q$ may be performed in time $O(n^2)$ by Lemma~\ref{lem:radial-sort}, after which the time to process each point $q$ is linear.
\end{proof}

\begin{theorem}
\label{thm:quadgraph-alg}
If a set $P$ of $n$ points has no empty pentagon, we can construct $\quadgraph P$ in time $O(n^2)$.
\end{theorem}

\begin{proof}
As in the algorithm for finding an empty pentagon, we rotate
the points so that no two have the same $y$-coordinate, sorting the points above each point $q$ radially around $q$, and triangulate the polygon formed by connecting the points in this radial order. This triangulation must be unique, for otherwise it would be possible to flip some two of its triangles, but then (as in the proof of Lemma~\ref{lem:test-bottom}) there would be an empty pentagon in~$P$.

A similar argument to that in Lemma~\ref{lem:test-bottom} shows that, in an input set with no empty pentagon, a set of four points forms the endpoints of the diagonals in a flip, with $q$ having the minimum $y$-coordinate of the four points, if and only if the other three points form one of the triangles in the triangulated polygon for $q$. In more detail, let $a$, $b$, and $c$ be any three points in radial order $abc$ as viewed from $q$.  In one direction, suppose some two of these three points are not visible to each other within the polygon. then $abcq$ cannot be an empty convex quadrilateral and does not represent a flip in $\quadgraph$; in this case, also, $abc$ cannot be a triangle in the triangulation.

In the other direction, suppose that all three points $a$, $b$, and $c$ are visible to each other. Then diagonal $ac$ must pass within the polygon, below point $b$, so $abcq$ is empty and convex and represents a flip in $\quadgraph$. In this case, by the visibility of its three vertices, $abc$ is a triangle in some triangulation of the polygon, and by the uniqueness of the triangulation it is a triangle in the constructed triangulation.

Thus, in total time $O(n^2)$ we may find the flips forming all of the edges in $\quadgraph$. Combining these edges to form an adjacency list representation of the graph $\quadgraph$ itself is straightforward within the same time bound.
\end{proof}

\begin{figure}[t]
\centering\includegraphics[width=6in]{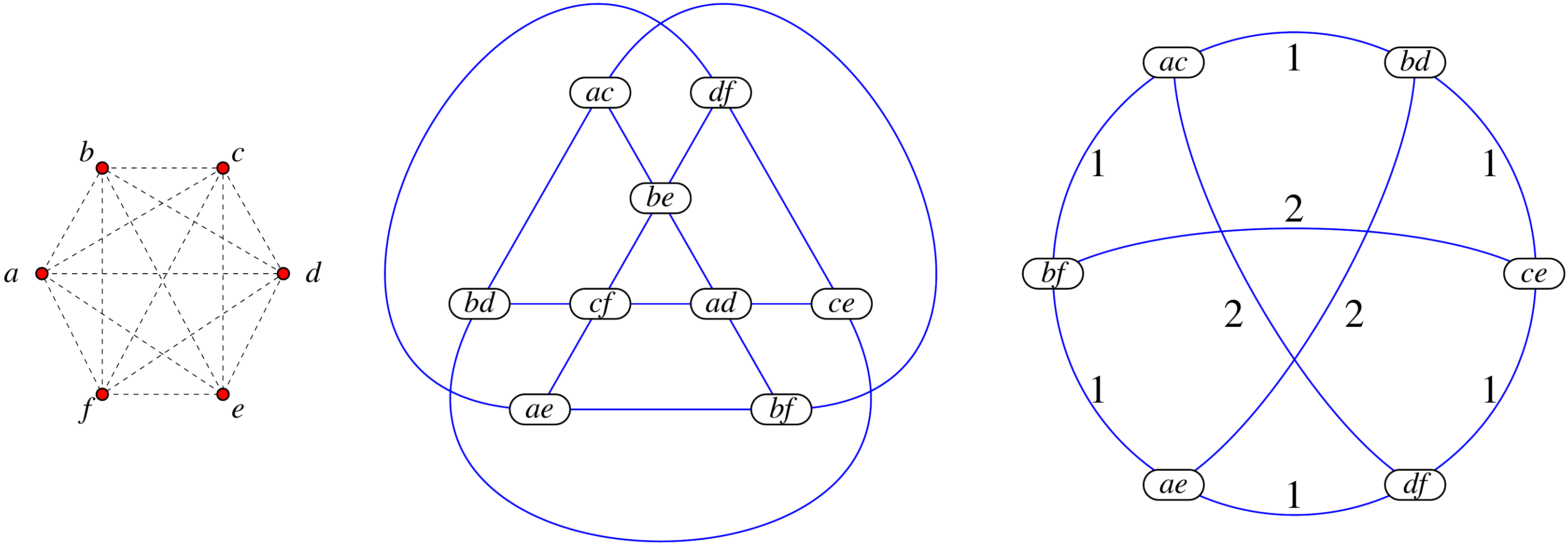}
\caption{A hexagon (left), its quadrilateral graph (center), and the complete bipartite graph of distances in the quadrilateral graph from $\{ac,ae,ce\}$ to $\{bd,bf,df\}$ (right).}
\label{fig:hexest}
\end{figure}

\section{Computing and estimating flip distance}

From the partial cube embedding in Theorem~\ref{thm:pent-equivalence} it follows that the flip distance between two triangulations $T$ and $T'$ in a point set $P$ with no empty pentagon may be computed by finding the edges $e$ from $T$ and $e'$ from $T'$ that correspond to vertices in each component of $\quadgraph P$, computing the distance between the corresponding pair of vertices in $\quadgraph P$, and adding the numbers obtained in this fashion for each component of $\quadgraph P$. We observe that this distance computation can be implemented as an algorithm that runs in time $O(n^2)$ without explicitly constructing $\quadgraph P$: flip both triangulations to the Delaunay triangulation, find the flips that are used in one but not both of these two flip sequences, and return the number of these flips.

We now describe a technique for estimating flip distances heuristically in any point set.

\begin{theorem}
Let $T$ and $T'$ be any two triangulations of the same point set $P$. Form a complete bipartite graph with the diagonals of $T$ on one side and the diagonals of $T'$ on the other, and label the edge between any two diagonals by the distance between those diagonals in $\quadgraph P$. Let $M$ be the minimum weight of a perfect matching in this complete bipartite graph. Then $M$ can be computed in polynomial time, and provides an underestimate of the true flip distance between $T$ and $T'$.
\end{theorem}

\begin{proof}
For any sequence of flips that converts one triangulation $T$ to another $T'$, we may partition the flips into a collection of paths in $\quadgraph P$, in which each path starts at a vertex of $\quadgraph P$ representing a diagonal of $T$ and ends at a vertex representing a diagonal of $T'$: at the start of the sequence of flips form a length-0 path for each diagonal in $T$, and whenever flipping a diagonal $xy$ to another diagonal $uv$, concatenate the quadrilateral graph edge $xy$-$uv$ onto the end of the path that previously ended in $xy$. At each stage of this procedure the current set of paths each start at a vertex representing a diagonal of the current triangulation, so when this procedure finishes processing all flips it will have formed the desired set of paths.

Thus, the total number of flips in the flip sequence must be equal to the total lengths of these paths.
The edges in the matching correspond to a set of paths in $\quadgraph P$ connecting the same endpoints, with minimum total length among all possible such sets of paths; therefore, the total weight of the matching is an underestimate of the length of any flip sequence.
\end{proof}

\begin{figure}[t]
\centering\includegraphics[width=4in]{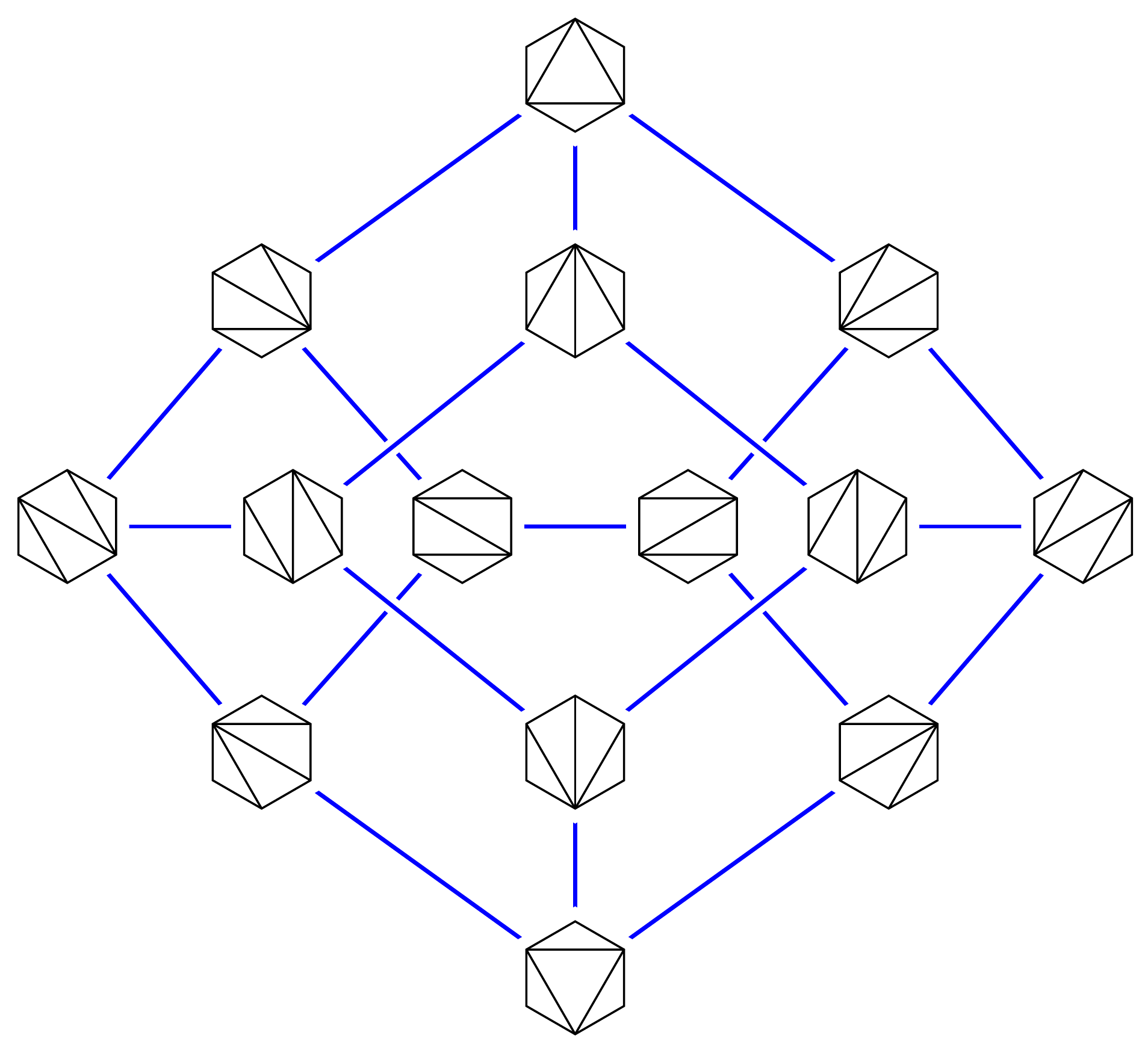}
\caption{The flip graph of a hexagon. Note that the top and bottom triangulations are at distance four from each other.}
\label{fig:fg6}
\end{figure}

For instance, consider the hexagon shown on the left of Figure~\ref{fig:hexest}, with $\quadgraph P$ shown in the center of the same figure.  There are two triangulations of the hexagon that use none of its long diagonals: one, $T$, using diagonals $\{ac,ae,ce\}$, and the other, $T'$, using diagonals $\{bd,bf,df\}$. On the right of Figure~\ref{fig:hexest} is a complete bipartite graph with the diagonals $\{ac,ae,ce\}$ on one side of the bipartition and the diagonals $\{bd,bf,df\}$ on the other side.  It is easily seen that the minimum weight matching in this complete bipartite graph has total weight three. Thus, we can conclude that any flip sequence transforming $T$ to $T'$ requires at least three flips. However, the actual flip distance between $T$ and $T'$ in this example is four, as can be seen from the flip graph of the hexagon (Figure~\ref{fig:fg6}).

This estimate may be useful in $A^*$-algorithm based heuristic search for the true flip distance between two triangulations, as $A^*$ depends on the existence of an estimate that is guaranteed to be an underestimate~\cite{HarNilRap-SSC-68}. Additionally, there are some point sets $P$ for which this method is guaranteed to produce the flip distance correctly rather than yielding an inaccurate estimate. For instance, if $P$ has no empty pentagons, then a pair of diagonals from $T$ and $T'$ will have finite distance in $\quadgraph P$ if and only if they belong to the same tree in $\quadgraph P$, so the minimum weight perfect matching will match each diagonal of $T$ with the unique diagonal of $T'$ in the same tree of $\quadgraph P$. Thus, in this case, our estimate ends up calculating the flip distance exactly.  If $P$ is itself a pentagon, also, this estimate can be seen by a simple case analysis to calculate the flip distance exactly. However, whenever $P$ contains an empty hexagon, there exist triangulations $T$ and $T'$ for which $M$ is a strict underestimate of the flip distance, because we can embed the two triangulations with diagonals $\{ac,ae,ce\}$ and $\{bd,bf,df\}$ from the example above into the empty hexagon of $P$. Thus, the nonexistence of an empty hexagon is a necessary condition for our estimate to be exact. It would be of interest to determine whether it is also a sufficient condition.

\raggedright
\bibliographystyle{abuser}
\bibliography{happy-end}

\end{document}